# Domain Walls and Anchoring Transitions Mimicking Nematic Biaxiality in the Oxadiazole Bent-Core Liquid Crystal C7

Young-Ki Kim,[a] Greta Cukrov,[a] Jie Xiang,[a] Sung-Tae Shin,[b] and Oleg D. Lavrentovich*[a]

[a] *Liquid Crystal Institute and Chemical Physics Interdisciplinary Program, Kent State University, Kent, OH 44242, USA*

[b] *Department of Chemical Engineering, Kyung Hee University, Youngin, Gyunggi-do 446-701, Republic of Korea*

* Corresponding Author. E-mail : olavrent@kent.edu

## Abstract

We investigate the origin of "secondary disclinations" that were recently described as a new evidence of a biaxial nematic phase in an oxadiazole bent-core thermotropic liquid crystal C7. With an assortment of optical techniques such as polarizing optical microscopy, LC PolScope, and fluorescence confocal polarizing microscopy, we demonstrate that the secondary disclinations represent non-singular domain walls formed in an uniaxial nematic during the surface anchoring transition, in which surface orientation of the director changes from tangential (parallel to the bounding plates) to tilted. Each domain wall separates two regions with the director tilted in opposite azimuthal directions. At the centre of the wall, the director remains parallel to the bonding plates. The domain walls can be easily removed by applying a modest electric field. The anchoring transition is explained by the balance of (a) the intrinsic perpendicular surface anchoring produced by the polyimide aligning layer and (b) tangential alignment caused by ionic impurities forming electric double layers. The model is supported by the fact that the temperature of the tangential-tilted anchoring transition decreases as the cell thickness increases and as the concentration of ionic species (added salt) increases. We also demonstrate that the surface alignment is strongly affected by thermal degradation of the samples. The study shows that C7 exhibits only a uniaxial nematic phase and demonstrate yet another mechanism (formation of "secondary disclinations") by which a uniaxial nematic can mimic a biaxial nematic behaviour.

# 1. Introduction

The search for the biaxial nematic ($N_b$) liquid crystals (LCs) represents an active and fascinating area of studies with arguments presented both in favor[1-25] and against[26-36] their existence. In the case of lyotropic surfactant systems, the discussion issue is stability.[1, 26, 27] Among the thermotropic LCs, the most studied is the oxadiazole bent-core material 4,4'(1,3,4-oxadiazole-2,5-diyl) di-*p*-heptylbenzoate derived from 2,5-bis-(*p*-hydroxyphenyl)-1,3,4-oxadiazole, abbreviated ODBP-Ph-$C_7$, or simply C7 (Figs. 1a,b). The $N_b$ phase of C7 has been suggested by X-ray diffraction (XRD),[6, 7] NMR,[8] and electro-optical studies.[9-12] These suggestions were challenged[28-32] on the grounds that the observed features represent a mimicking behaviour of the uniaxial nematic ($N_u$) phase rather than a true biaxial order in the bulk of the specimen. For example, the $N_u$ phase of C7 might mimic a biaxial behaviour through the appearance of smectic cybotactic groups,[29, 30] and surface anchoring transitions in which the uniaxial director $\hat{\mathbf{n}}$ (depicting the average direction of long molecular axes) tilts as the sample's temperature is varied.[31, 32]

Recently, a new evidence has been presented in favor of the $N_b$ phase of C7, based on the polarizing optical microscopy (POM) textures exhibiting the so-called "secondary disclinations" (SDs).[37] The POM showed a secondary Schlieren texture growing into a $N_u$ texture on lowering the temperature and disappearing on heating.[37] The appearance of SDs was attributed to the $N_u – N_b$ phase transition. The SDs were associated with disclinations in the orientational order of the short molecular axes, i.e., with the secondary director (different from $\hat{\mathbf{n}}$) of the $N_b$ phase.[37]

In this paper, we apply a battery of imaging techniques, namely, POM, LC PolScope, and fluorescence polarizing optical microscopy (FCPM), to explore the C7 textures and their temperature and electric field-induced behaviour in cells with different surface alignment layers. Our studies demonstrate that the SDs are caused by a surface anchoring transition, i.e., by realignment of the uniaxial director $\hat{\mathbf{n}}$ from being parallel to the bounding plates to being tilted. The SDs thus represent domain walls (DWs) that separates regions of tilted director with different azimuthal direction. The anchoring transition is associated with the balance of two different mechanisms responsible for the alignment of C7: alignment by the polyimide substrate and by the electric field within the surface electric double layer formed by ionic impurities. The findings demonstrate a uniaxial nature of the nematic order in C7 and are thus consistent with the previous claims of uniaxial character of C7.[31, 32] The DWs observed in C7 are similar to the DWs observed in another uniaxial nematic LC, with the H-shaped molecules.[38]

## 2. Materials and Techniques

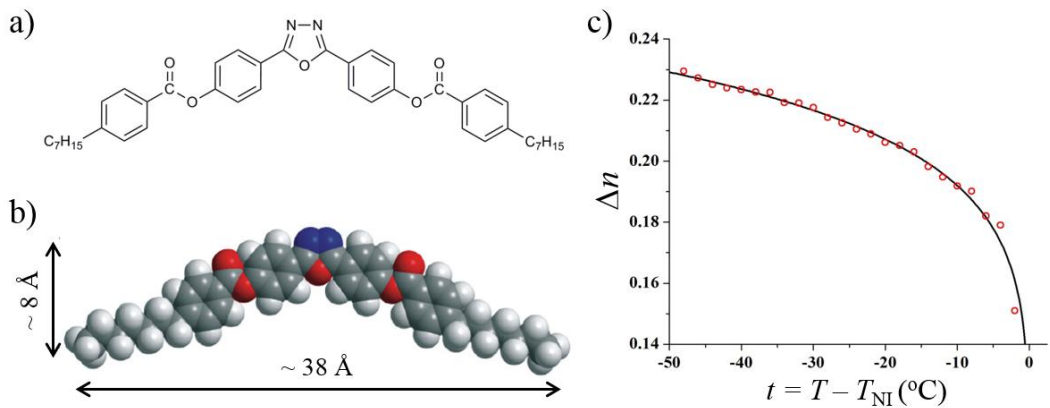

**Fig. 1** (a) Chemical structure and (b) space filling model of the oxadiazole bent-core liquid crystal C7. (c) Birefringence $\Delta n$ of C7 as a function of the relative temperature $t = T - T_{NI}$.

Figs. 1a,b show the chemical structure of the oxadiazole bent-core LC, C7. The phase diagram confirmed by a POM upon cooling is

$$\text{Cr } (148\,^\circ\text{C}) \text{ Sm } (173\,^\circ\text{C}) \text{ N } (222\,^\circ\text{C}) \text{ I}$$

where Cr, Sm, N, and I are crystal, smectic, nematic, and isotropic phase, respectively; the phase diagram is consistent with the previous studies,[7, 8, 31] if one does not discriminate between the two nematic ($N_u$ and $N_b$) phases.

The cells were assembled from flat glass substrates with alignment layers. The cell thickness $d$ was set by glass spacers mixed with UV glue NOA 65 (Norland Products, INC.) that was also used to seal the cells. Because C7 shows signs of aging at elevated temperatures in presence of oxygen,[31] we performed the experiments with freshly prepared cells within 5 hours or less. In order to prevent possible surface memory effect,[31] we injected C7 into the cells at the temperatures above the N-I transition temperature $T_{NI}$ and carried out the experiments within the temperature range of the N phase. The temperature was controlled by a LTS420 hot stage with a T95-HS controller (both Linkam Instruments), with $0.01\,^{\circ}$C accuracy. The rate of temperature change $\xi = \pm 0.4\,^{\circ}$C/min was deliberately slow to minimize the flows caused by thermal expansions that might represent yet another facet of a uniaxial behaviour that mimics the behaviour of $N_b$.[35, 39, 40]

In the analysis of optical textures, it is important to know the temperature dependence of birefringence $\Delta n$. The latter is presented in Fig. 1c as a function of the relative temperature $t = T - T_{NI}$; the dependence was measured in planar cells of thickness $d = 1.1\,\mu\text{m}$ with polyimide PI2555 (HD Microsystems) serving as planar alignment layers.

## 3. Alignment, Anchoring Transition and Domain Walls

To re-enact the appearance of SDs, we fabricated cells with two polyimide aligning agents, SE1211 and SE5661 (both supplied by Nissan Chemical Industries) deposited onto the indium-tin-oxide (ITO) electrodes at glass substrates. The polyimide layers were spin-coated from solutions and then baked but not rubbed.

### 3.1 Polarizing Optical Microscopy

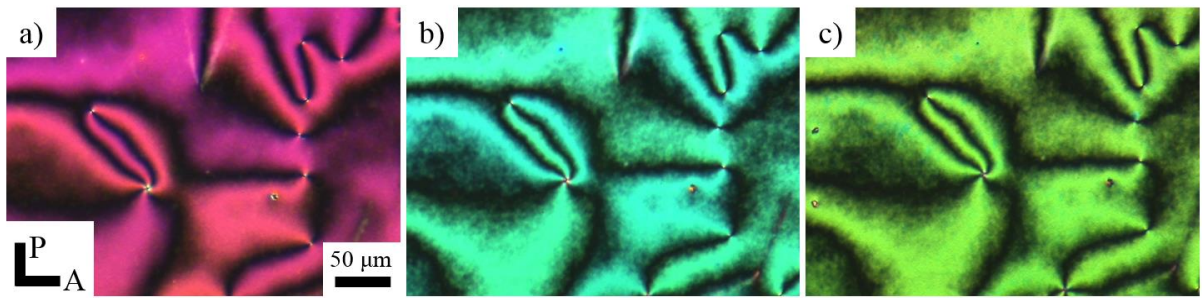

**Fig. 2** Schlieren textures of C7 in the SE1211 cell ($d = 4.5\,\mu\text{m}$) at (a) $t = -1.0$, (b) $-20.0$, and (c) $-45.0\,^{\circ}$C observed under a POM. A and P are crossed analyzer and polarizer directions, respectively.

In the SE1211 cells, one observes the classic Schlieren textures characteristic of the uniaxial order in the entire temperature range of the N phase, Fig. 2.[41] The Schlieren textures feature point defect-boojums (with four brushes of extinction emerging from their centers) and vertical disclinations of semi-integer strength (with two dark brushes emanating from their cores), Fig. 2.[42]

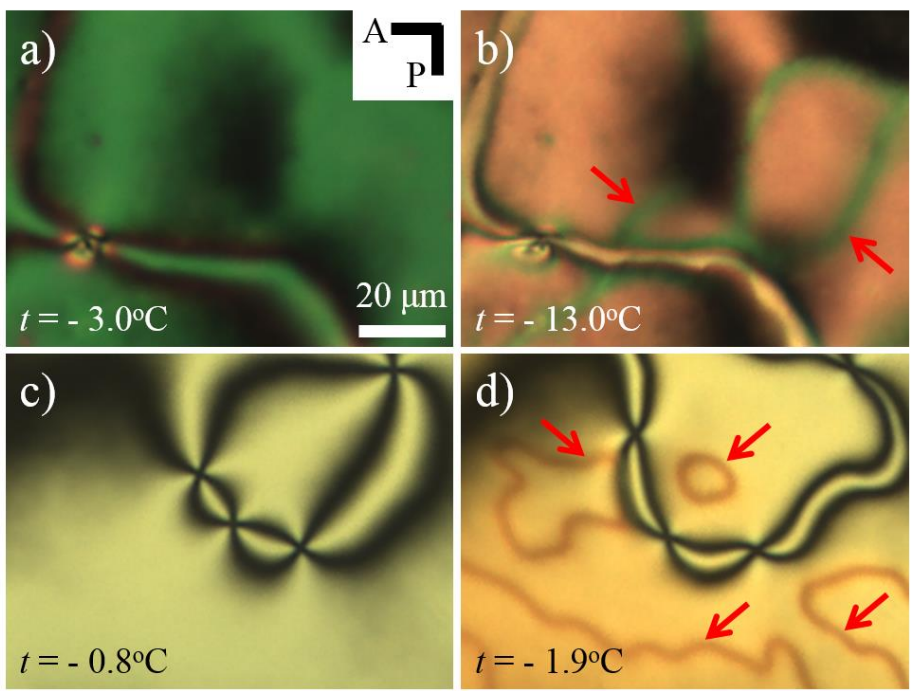

**Fig. 3** Temperature dependence of POM textures in the SE5661 cells of different thickness: (a, b) $d = 13\,\mu m$ ($t^* = -5.5\,°C$) and (d, f) $4.5\,\mu m$ ($t^* = -0.9\,°C$). The left column shows standard Schlieren textures at elevated temperatures $t > t^*$, and the right column shows well-defined SDs at $t < t^*$. Red arrows point towards the SDs.

In the SE5661 cells, one observes two different textures as the temperature is varied. Above some critical temperature $t^*$, the textures are of the classic $N_u$ Schlieren type, Figs. 3a,c. Below $t^*$, secondary textures with SDs form, Figs. 3b,d. The SDs show a reversible behaviour, disappearing once the temperature is raised above $t^*$ and re-appearing again when the sample is cooled below $t^*$. The transition temperature $t^*$ increases as the cell thickness $d$ decreases, Fig.

3; $t^* = -5.5\,°C$ when $d = 13\,\mu m$, and $t^* = -0.9\,°C$ when $d = 4.5\,\mu m$. In very thin cell, $d = 1.1\,\mu m$, SDs appear simultaneously with the formation of the N phase; $t^* \approx 0\,°C$. These thin cells are convenient for the quantitative analysis of the textures through LC PolScope observations that map the pattern of optical retardance of the cell, Fig. 4, as described in the next section.

### 3.2 Maps of retardance and director field by LC Polscope observation

The LC PolScope (Abrio Imaging System) allows one to map the optical retardance $\Gamma = \Delta n_{eff} d$ of the studied cells as the function of in-plane coordinates $(x, y)$, where

$$\Delta n_{eff} = \frac{n_o n_e}{\sqrt{n_e^2 \sin^2\theta + n_o^2 \cos^2\theta}} - n_o \qquad (1)$$

is the effective birefringence determined by the ordinary $n_o$ and extraordinary $n_e$ refractive indices; $\theta$ is the angle between $\hat{\mathbf{n}}$ and the bounding plate. When $\theta = 0$, the optical retardance reaches its maximum, $\Gamma_{max} = \Delta n d$, since in this case $\Delta n_{eff} = \Delta n = n_e - n_o$. In addition to mapping $\Gamma(x, y)$, LC PolScope also shows the local projection of the optic axis $\hat{\mathbf{n}}$ onto the cell's plane, Fig. 4.[43-45]

Figs. 4a-d show the temperature dependence of LC PolScope textures in the SE5661 cell of $d = 1.1\,\mu m$. The SDs exist in the entire temperature range of the N phase ($t^* \approx 0\,°C$), manifesting themselves as bands of high retardance $\Gamma_{max}$, separating regions of lower $\Gamma < \Gamma_{max}$. As $t$ decreases, $\Gamma$ measured in the surrounding domains decreases, Fig. 4e. Such a behaviour is consistent with the idea that the SDs are domain walls (DWs) separating regions with the titled $\hat{\mathbf{n}}$, $\theta > 0$. At the center of the DW, $\hat{\mathbf{n}}$ has to be parallel to the plates, $\theta = 0$, in order to connect the two regions with different azimuthal directions of the tilt $\theta > 0$, Fig. 5. Fig. 5 shows the details of director field in the DWs that can be associated with splay-bend (Fig. 5b) and twist (Fig. 5c) deformations, depending on how the DW is oriented with respect to the projection of $\hat{\mathbf{n}}$ onto the cell's substrates.

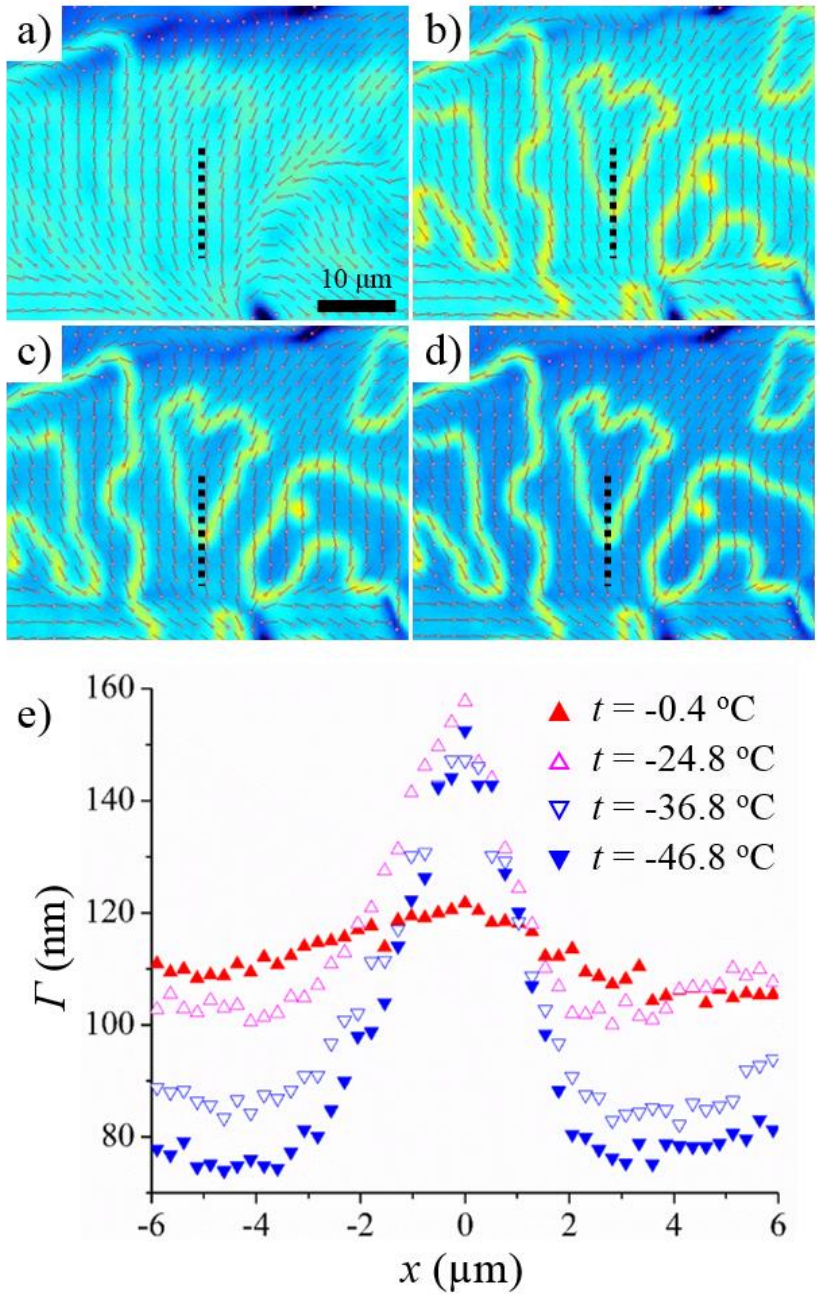

**Fig. 4** Retardation maps of SD textures in the SE5661 cell ($d = 1.1\,\mu\text{m}$) at (a) $t = -0.4\,(<t^* \approx 0\,^\circ\text{C})$, (b) $-24.8$, (c) $-36.8$, and (d) $-46.8\,^\circ\text{C}$; ticks show the projection of $\hat{\mathbf{n}}$; the central parts of the SDs feature horizontally aligned $\hat{\mathbf{n}}$ and thus yield a higher value of $\Gamma$. The SDs separate regions with tilted $\hat{\mathbf{n}}$ and thus reduced $\Gamma$. (e) Variation of $\Gamma$ across the SD along the pathway shown by dotted lines in (a-d).

Outside the DW, the director tilt $\theta$ increases as the temperature is lowered below $t^*$, thus both $\Delta n_{eff}$ and $\Gamma$ decrease, according to Eq. (1), see Fig. 4e. At the center of DW, $\hat{\mathbf{n}}$ remains parallel to the bounding plates before and after the anchoring transition. One would expect the measured retardance to reach its maximum value $\Gamma_{\max} = \Delta nd$ there. Although the $\Gamma$ peaks at the DW, Fig. 4e, limited lateral resolution of the LC PolScope (being of the same order as the width of the DWs, about $1\,\mu\mathrm{m}$) does not allow one to map the $\Gamma$ within the DWs accurately; the value of $\Gamma$ at the center of the DWs is somewhat decreased as compared to the expected value $\Gamma_{\max} = \Delta nd$ because of the surrounding regions with the tilted director. However, this problem of limited in-plane resolution is eliminated in the experiments with the electric field, as described below.

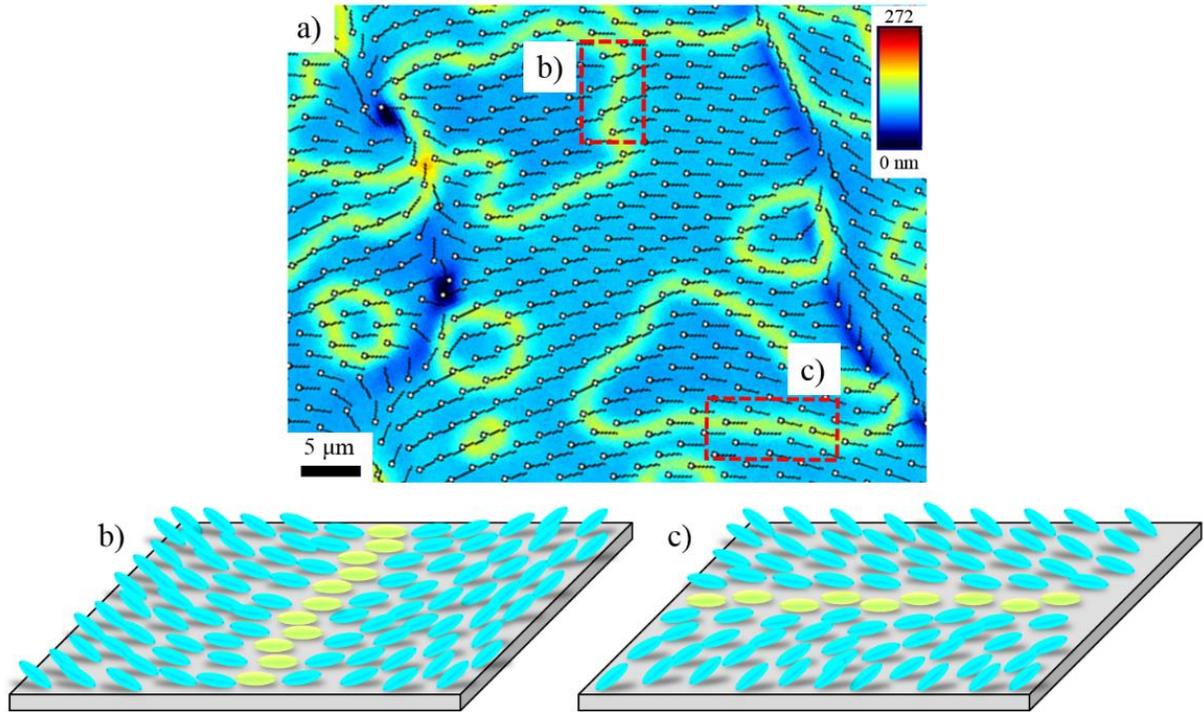

**Fig. 5** Retardation map with the director field projected on the plane of the SE5661 cell ($d = 1.1\,\mu\mathrm{m}$) at $t = -46.8\ ^\circ\mathrm{C}\,(< t^*)$. Director configurations of DWs associated with (b) bend-splay and (c) twist deformations.

## 3.3 Behaviour of textures in the electric field

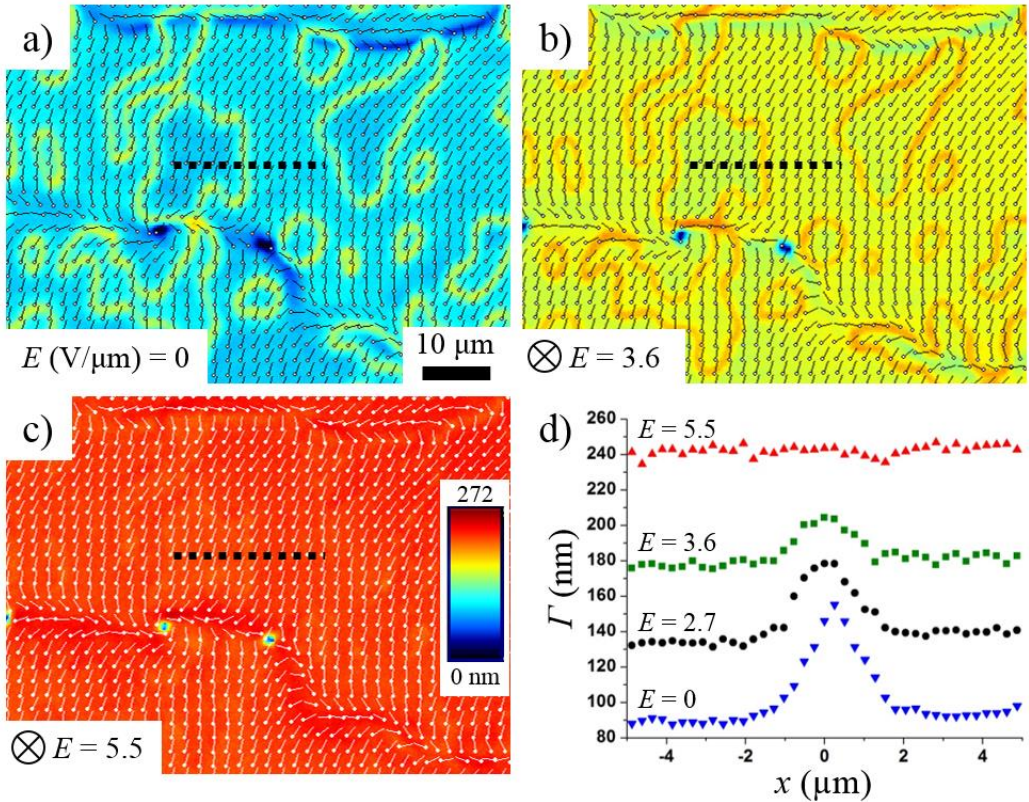

**Fig. 6** Retardation maps of the SE5661 cell ($d = 1.1\,\mu m$) at $t = -36.8\,^{\circ}\text{C}$ ($< t^*$) under the action of the vertical electric field; (a) $E = 0$, (b) 3.6, and (c) 5.5 $V_{rms}/\mu m$. (d) Change of $\Gamma$ across a DW (along dotted lines) as a function of $E$.

In order to further confirm the model of DWs, we performed an experiment in which an alternating current (AC) field $E$ (sinusoidal wave of frequency $f = 1\,\text{kHz}$) was applied across the cell, between two ITO electrodes on the glass plates, Fig. 6. C7 has a negative anisotropy of dielectric permittivity: $\Delta\varepsilon\,(t = -2\,^{\circ}\text{C}) = -17.1$ at $f = 1\,\text{kHz}$,[31] which means that the field realigns $\hat{\mathbf{n}}$ perpendicularly to itself. In absence of $E$, the retardation $\Gamma$ outside the DWs is very small, Fig. 6a. An applied $E$ increases $\Gamma$ in these homogeneous regions, Fig. 6b, up to the point (achieved at $E = 5.5\,V_{rms}/\mu m$) when $\Gamma$ of the entire sample is uniformly high and the DWs are no longer distinguishable (Figs. 6c,d). This behaviour is consistent with the model of DWs separating differently tilted domains. Namely, the electric field forces the tilted $\hat{\mathbf{n}}$ to realign everywhere parallel to the bounding plates, thus eliminating the DWs.

The tilt angle $\theta$ achieved as the result of surface anchoring transition below $t^*$ can be estimated from the data on $t$ and Eq.(1), by using the values $n_e \approx \bar{n} + 2\Delta n/3$ and $n_o \approx \bar{n} - \Delta n/3$ determined by the measured $\Delta n$ (Fig. 1c) and the approximate value of the averaged refractive index $\bar{n} \approx 1.60$.[31, 46] At $t = -36.8\ °\text{C}$, with the estimated $n_e \approx 1.75$ and $n_o \approx 1.53$, one finds that $\theta(E=0) \approx 50°$ (Fig. 6a), $\theta(E = 3.6\ \text{V/µm}) \approx 28°$ (Fig. 6b), and $\theta(E = 5.5\ \text{V/µm}) < 1°$ (Fig. 6c) in the SE5661 cell of $d = 1.1\ \text{µm}$. Figs. 6c,d show that $\Gamma$ of the sample in the electric field of amplitude $5.5\ \text{V}_{rms}/\text{µm}$ is about $243\ \text{nm}$. Since $d = 1.1\ \text{µm}$, this result implies that the quantity $\Gamma/d = 0.22$ is in an excellent agreement with the value of $\Delta n$ measured independently in the planar well aligned field-free cells at $t = -36.8\ °\text{C}$, Fig. 1c.

On the basis of the POM and LC PolScope studies, we conclude that the SDs are not associated with the appearance of the secondary director and represent DWs that emerge during a temperature-induced surface anchoring transition in the uniaxial nematic cells with certain types of surface aligning layers (such as SE5661). Below $t^*$, the director $\hat{\mathbf{n}}$ deviates from the tangential alignment; the regions with different azimuthal directions of the tilt are bridged by the DWs. The tilted orientation of $\hat{\mathbf{n}}$ below $t^*$ is also confirmed by the fact that in POM and LC PolScope textures, the DWs either join two half integer disclinations or form closed loops (Fig. 3). Below $t^*$, the projection of the tiled $\hat{\mathbf{n}}$ onto the plane of the cell is a vector (as opposed to the director), thus the half-integer disclinations no longer exist as isolated defects and must be associated with the DWs.[47] The fact that $t^*$ depends on $d$ (Figs. 3 and 4) provides another evidence that the DWs are associated with the effects of confinement rather than with the $\text{N}_u - \text{N}_b$ phase transition in the bulk; the thermodynamic stability of the biaxial nematic $\text{N}_b$ should not depend on the type of surface alignment and on variations of $d$ in the range of micrometers.

### 3.4 Fluorescence confocal polarizing microscopy of the surface anchoring transition

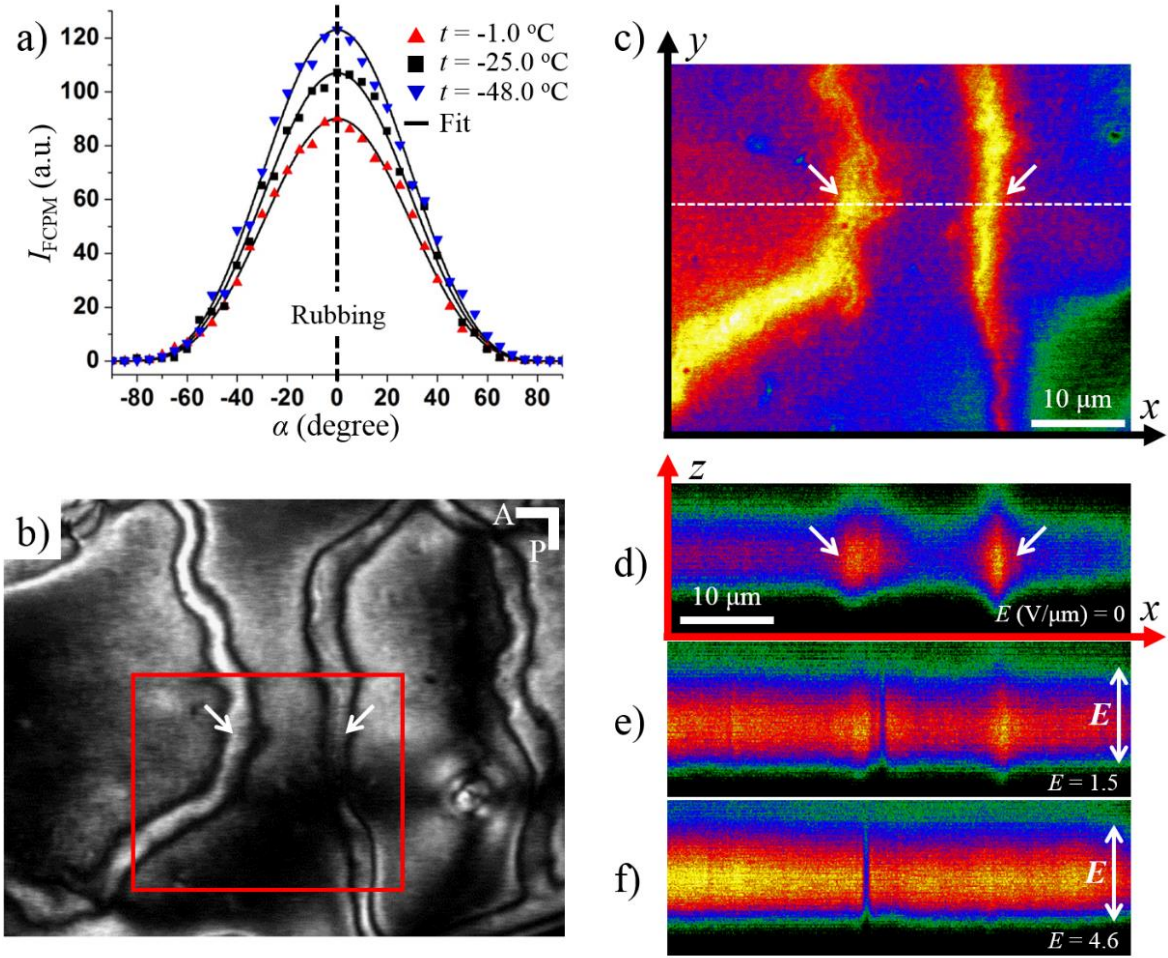

**Fig. 7** (a) FCPM intensity $I_{FCPM}$ in the rubbed planar cell (PI2555) of C7 as a function of the angle $\alpha$ and the temperature $t$. (b) POM texture of the SE5661 cell ($d = 13\,\mu m$) between crossed polarizers at $t = -37.0\,^{\circ}C\,(<t^*)$; arrows indicate DWs. FCPM textures scanned in (c) the horizontal $(x, y)$ plan at the squared region in b) and (d-f) the vertical $(x, z)$ plan along dotted line in c) under the action of the vertical electric field; (d) $E = 0$, (e) $1.5$, and (f) $4.6\,V_{rms}/\mu m$; arrows indicate DWs.

For the direct demonstration of the anchoring transition and director tilt at $t < t^*$, the SE5661 cell was investigated using the 3D microscopy technique, so-called fluorescence confocal polarizing microscopy (FCPM).[48] C7 was doped with a small amount ($0.01\,wt\%$) of the fluorescence dye N,N'-Bis(2,5-di-*tert*-butylphenyl)-3,4,9,10-perylenedicarboximide (BTBP, Sigma Aldrich). Planar (PI2555) cells were used to establish that the transition dipole $\hat{\mathbf{d}}$ of BTBP molecules is parallel to $\hat{\mathbf{n}}$ of C7 by measuring the azimuthal angular dependence of the fluorescence intensity $I_{FCPM}$ (Fig. 7a).[48, 49] $I_{FCPM}$ reaches its maximum when the angle $\alpha$ between $\hat{\mathbf{n}}$ and the polarization of probing light is zero; $I_{FCPM} \propto \cos^4 \alpha$.[48, 49] FCPM textures of PI2555 cells show that $I_{FCPM}$ does not change when the electric field is applied across the cell, confirming strict tangential alignment of $\hat{\mathbf{n}}$.

FCPM shows an evidence of an anchoring transition in the SE5661 cells, Figs. 7c-f. At $t = -37.0\ ^\circ\text{C}$, the SE5661 cell of thickness $d = 13\ \mu\text{m}$ viewed by a POM with two crossed polarizers shows the DWs (indicated by arrows in Fig. 7b). The same region of the sample was scanned in the FCPM mode, in the plane parallel to the bounding plates, Fig. 7c, and in the plane perpendicular to the bounding plates, Figs. 7d-f. The probing beam was circularly polarized in order to detect only the polar angle of the director tilt. In both the horizontal $(x, y)$ scans (Fig. 7c) and the vertical $(x, z)$ scans (Fig. 7d), the DWs feature a higher intensity of fluorescence as compared to that of the surrounding regions. This feature is consistent with the idea that $\hat{\mathbf{n}}$ in the center of DWs is parallel to the bounding plates and is tilted in the regions outside the DWs. Furthermore, the vertical scans obtained in the presence of $E$ (Figs. 7e,f) show that $I_{FCPM}$ increases as $E$ is increased. This clearly demonstrates that the electric field realigns $\hat{\mathbf{n}}$ towards the horizontal planes (parallel to the bounding glass plates).

### 3.5 Thickness-dependent anchoring transition and electric double layers of ions

The appearance of anchoring transition at $t^*$ for some aligning substrates and the fact that $t^*$ depends on the cell thickness can be explained by the aligning action of electric double layers formed by ionic impurities near the substrates. The electric double layers create local electric fields acting on the director near the surfaces.[50, 51] The surface anchoring potential for a tangentially anchored substrate is, in Rapini-Papoular approximation,

$$W(\theta) = W \sin^2 \theta, \tag{2}$$

where the positive definite anchoring coefficient $W = W_0 + W_i > 0$ can be represented as a sum of an "intrinsic" coefficient $W_0 < 0$ and an "electrostatic" $W_i > 0$ contributions, caused by the dielectric torque of the local electric field on the dielectrically anisotropic LC. The latter was estimated by Barbero and Durand[50, 52] to be

$$W_i = -\frac{\sigma^2 \Delta \varepsilon \lambda_D}{2\varepsilon_0 \varepsilon^2}, \qquad (3)$$

where $\sigma$ is surface density of charges, $\varepsilon$ is the average dielectric permittivity of the LC, $\varepsilon_0$ is the electric constant, and $\lambda_D$ is the thickness of the double layers, also known as the Debye screening length (see also Ref. 53 for the range of validity of Eq. (3)). The local electric field is perpendicular to the bounding surfaces. Therefore, since C7 is of a negative dielectric anisotropy, $\Delta\varepsilon < 0$, the local field tends to align the director parallel to the surface, so that the electrostatic coefficient $W_i > 0$ facilitates tangential anchoring. The intrinsic anchoring at polyimides such as SE5661 is typically homeotropic, $W_0 < 0$. This anchoring is a local property of the interface, and is thus thickness – independent. The electrostatic part $W_i > 0$, however, is thickness dependent because the surface density $\sigma$ of absorbed ions at the interface depends on the total volume of the LC, so that $\sigma = \sigma_0 d / (d + 2\lambda_D)$; the value of $\sigma_0$ depends on availability of surface absorption sites and adsorption energy.[52] In LCs, $\lambda_D$ is relatively large, on the order of $0.1\text{-}1\,\mu\text{m}$,[54-56] and thus comparable to the range of cell thicknesses explored in this work. The electrostatic contribution to the surface anchoring is expected to reach its maximum positive value $W_i = \sigma^2(-\Delta\varepsilon)\lambda_D / 2\varepsilon_0 \varepsilon^2$ in thick samples, $d \gg \lambda_D$. As the cells become thinner, the anchoring coefficient decreases $W_i = \left(\sigma_0^2(-\Delta\varepsilon)\lambda_D / 2\varepsilon_0 \varepsilon^2\right)\left(d/(d+2\lambda_D)\right)^2$, which implies that the anchoring strength responsible for tangential alignment is weaker in thinner cells. This behaviour is in a qualitative agreement with our experiments, in which the thicker cells show the widest temperature range of the stable tangential alignment, and the thin cells show a very narrow or non-existent region of tangential alignment.

To further demonstrate the role of ions on the thickness-dependent anchoring transition, we performed the following two experiments. First, the transition temperature $t^*$ was measured as a function of concentration $c$ of the salt tetrabutylammonium bromide (TBAB, Sigma Aldrich),

                                                                                13

added to C7. As shown in Table 1, $t^*$ drops significantly with the increase of $c$, as expected. At higher $c$, the electric field of the double layers becomes stronger and imposes a stronger tangential alignment force. For high salt concentration $c = 2.9$ wt% and in thick cells, $d = 4.5\,\mu m$, there is no anchoring transition (and thus no DWs), as the tangential alignment persists in the entire range of the nematic phase.

**Table 1** Dependence of an anchoring transition temperature $t^*$ on the concentration $c$ of TBAB in the SE5661 cells of $d = 1.1$ and $4.5\,\mu m$.

|  | $c = 0$ wt% | $c = 1.0$ wt% | $c = 2.9$ wt% |
| --- | --- | --- | --- |
| $t^*\,(d = 4.5\,\mu m)$ | $-0.9\,^\circ C$ | $-31.1\,^\circ C$ | No Transition |
| $t^*\,(d = 1.1\,\mu m)$ | $0\,^\circ C$ | $-11.0\,^\circ C$ | $-24.0\,^\circ C$ |

The second experiment was designed to create an in-plane gradients of the ionic additives by applying the electric field in the plane of the cell. The SE5661 layer was spin-coated on a top glass substrate (no electrodes) and a bottom glass substrate with interdigitated ITO electrodes; the gap between electrodes was $1\,mm$. Two substrates were rubbed in the anti-parallel fashion and subsequently were assembled with $d = 1.3\,\mu m$; the rubbing direction is perpendicular to an in-plane field. Figs. 8a-c show the retardance maps of the cell at $t = -46.5\,^\circ C\,(<t^*)$ for different values of the in-plane electric field. In absence of the field, there is no significant variation of $\Gamma$, Figs. 8a, d. Once the in-plane direct current (DC) field is applied, $\Gamma$ measured in the cathode region increases, as compared to the rest of the cell, (Figs. 8b,d), and the higher value of $\Gamma$ on the cathode is maintained even after the field is removed (Figs. 8c,d). The in-plane DC field carries the negative ions to the cathode thus enhancing the tangential alignment, as evidenced by a higher value of $\Gamma$.

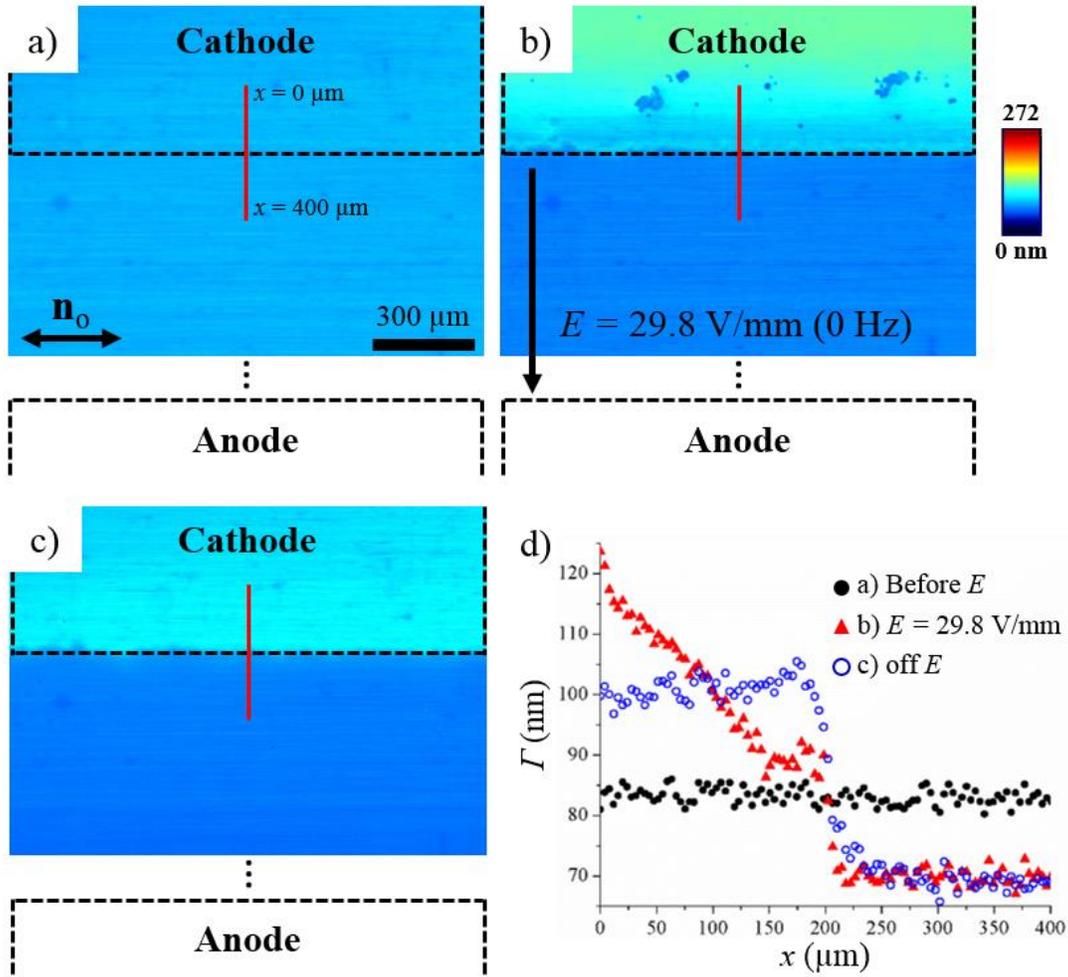

**Fig. 8** Retardation maps of the rubbed SE5661 cell ($d = 1.3\,\mu\text{m}$) at $t = -46.5\,^{\circ}\text{C}$ ($< t^*$) under the action of the in-plane electric field; (a) $E = 0$, (b) 30 min after $E = 29.8$ V/mm, and (c) 30 minutes after the field is off. (d) $\Gamma$ profile across an electrode (along red solid lines) in Figs. 8a-c.

Interestingly, a clear trend of thickness–dependent anchoring with the tangential contribution becoming weaker in thin cells has been recently observed by Ataalla, Barbero, and Komitov for the LC, MLC-6608, with $\Delta\varepsilon < 0$ in cells treated with SE1211.[57] The authors observed that the anchoring strength of homeotropic alignment increases as the cells become thinner, which is equivalent to the weakening of the electrostatically induced tangential anchoring $W_i$ at small $d$. The effect is similar to the phenomenon of thickness dependent $t^*$ for SE5661 cells observed in our work

## 3.6 Influence of thermal degradation of C7 on the alignment

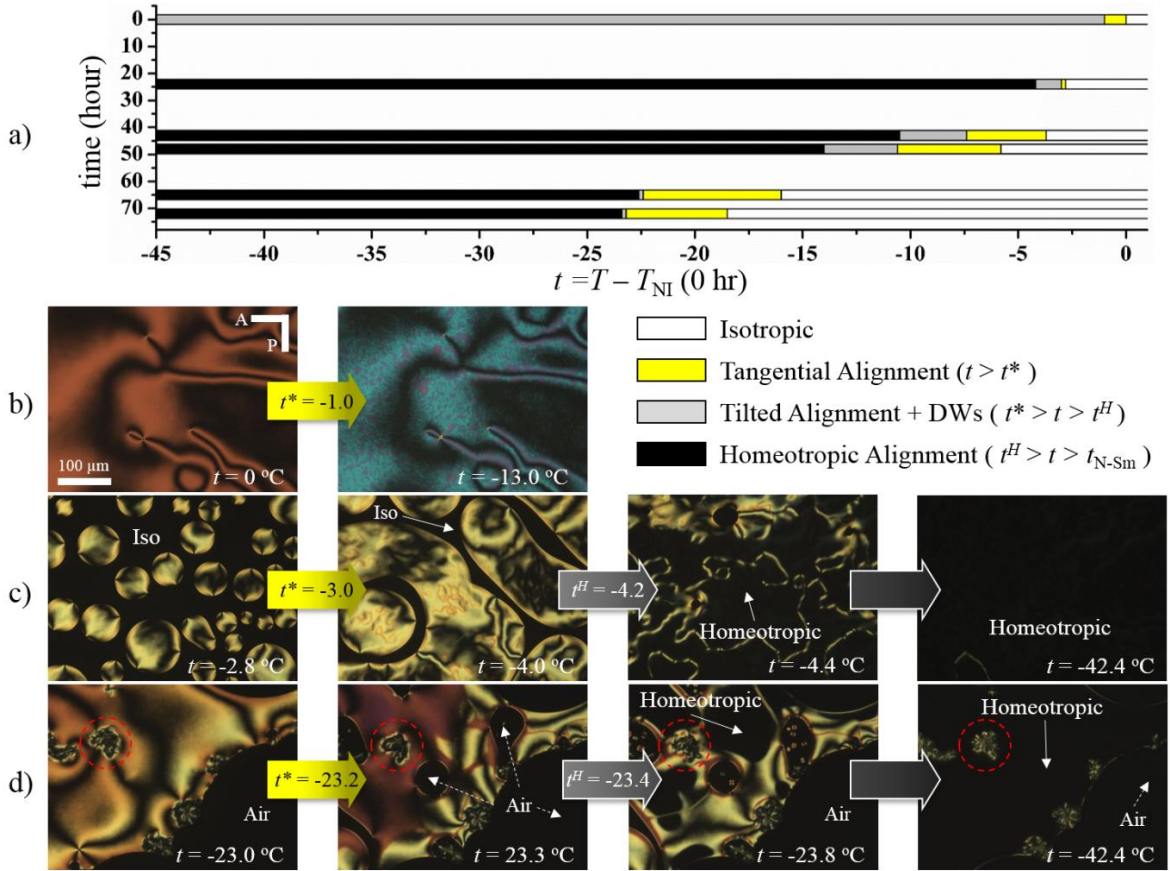

**Fig. 9** (a) Alignment dependence on thermal degradation of C7 in the SE5661 cells of $d = 4.5\,\mu m$. Temperature dependence of POM textures b) 0, c) 24, d) 72 hours after the cell was kept at $T = 220\,^\circ C$. The first column shows the POM textures of tangential alignment at $t > t^*$, the second column shows the POM textures of tilted alignment with DWs at $t^* > t > t^H$, the third column shows emerging homeotropic domains at $t^H > t$, and the forth column shows the cell entirely covered by homeotropic domains at $t = -42.4\,^\circ C\,(< t^H)$. Red dotted circle in (d) indicates the crystal aggregates.

It was reported that the C7 material is chemically unstable and experiences chemical degradation / decomposition when it is kept at the temperature of N phase ($222\,^\circ C > T > 173\,^\circ C$) in an oxygen environment.[31] This is the reason why all our experiments were performed with fresh cells within 5 hours or less.

To verify how the chemical degradation of C7 influences the alignment features, the director alignment in a SE5661 cell of $d = 4.5\,\mu m$ was observed over time while keeping the temperature fixed at $T = 220\,^\circ C$. The fresh cells exhibit the phase diagram as in the previous studies[7, 8, 31], and show an anchoring transition from tangential (1st column in Fig. 9b) to tilted alignment with the emergence of DWs at $t^*$ (2nd column in Fig. 9b). In the cells kept at elevated temperature for 24 hours, both $T_{NI}$ and the temperature range of tangential alignment decrease. Moreover, one observes another textural transition, from a tilted to homeotropic alignment state, at a certain temperature $t^H$, Figs. 9c. After a more prolonged exposure to the high temperature, 42-72 hours, the temperature range of tangential alignment expands, while $T_{NI}$ and the range of tilted alignment decrease, Fig. 9a. We also observed formation of air bubbles and crystal aggregates (red dotted circle in Fig. 9d) that did not melt even at $T = 270\,^\circ C$.

Therefore, the surface alignment of C7 is strongly affected by degradation at elevated temperatures.

## 4. Conclusions

In this study, we demonstrate that the so-called secondary disclinations (SDs) observed in the nematic C7 represent domain walls (DWs) that occur in a uniaxial nematic phase as a result of the surface anchoring transition triggered by temperature changes. This transition is observed for a certain aligning material (SE5661) but not for other aligning materials (such as PI2555 and SE1211). In the cells with the SE5661 aligning layer, above some temperature $t^*$, the director $\hat{\mathbf{n}}$ is parallel to the bounding plates. Below $t^*$, the director tilts away from the substrates. Directors in different regions of the cell tilt into different azimuthal directions. Topologically, these regions have to be bridged by the DWs in which the director remains tangential, Fig. 5.

The observed DWs are not associated with the biaxial nematic phase, as follows from the facts that (a) the DWs are observed only with some alignment layers and not with the others; (b) the temperature range of stability of DWs (its upper limit $t^*$) depends on the thickness of cells;

(c) the DWs can be removed by modest electric fields that realign the director from tilted orientation towards an orientation parallel to the bounding plates.

The issue of temperature-induced anchoring transitions in bent core mesogens is by itself an interesting problem. These transitions have been observed in C7 and C12, [31, 32] as well in some other materials.[32-34] Very interesting is the fact that the transition temperature $t^*$ depends on the cell thickness. The natural reason for thickness-dependent anchoring transitions is the presence of ionic impurities in the samples that form electric double layers and thus create local electric fields acting on the director.[50, 51] Since the dielectric anisotropy of C7 is negative, the vertical electric field of the double layers tends to align the director tangentially; this tendency competes with the perpendicular alignment caused by the polyimide layers. Additional experiments with variable concentration of added salts and in-plane gradients of the ions support the model of the anchoring transition as a balance of polyimide and electric double layers alignment tendencies. We also demonstrated that degradation of C7 at the temperatures corresponding to its nematic phase dramatically influence the surface anchoring of the material.

We conclude that C7 represents a uniaxial nematic phase in the entire temperature region between the isotropic melt and the smectic phase. The conclusion confirms an earlier statement on the uniaxial nature of C7 based on the studies of topological defects [31] and a magneto-optical response.[32] The study demonstrates yet another facet by which a regular uniaxial nematic mimics the properties of a biaxial nematic phase, this time through thickness dependent anchoring transition from a tangential to tilted alignment of the director.

**Acknowledgements**

We thank T. J. Dingemans, O. Francescangeli, L. A. Madsen, S. J. Picken, and E. T. Samulski for fruitful discussions. We thank V. Borsch, S. Shiyanovskii, B. Senyuk, and S. Zhou for valuable suggestions and assistances with experiments. The work is supported by NSF DMR grants 1121288.